\documentclass[prl,nofootinbib,floats,aps,superscriptaddress,twocolumn]{revtex4}

\usepackage{hyperref}
\usepackage{graphicx}
\usepackage{slashed}
\usepackage{amsmath,amssymb}
\usepackage{hhline}

\newcommand{\beq}{\begin{equation}}
\newcommand{\eeq}{\end{equation}}
\newcommand{\bea}{\begin{eqnarray}}
\newcommand{\eea}{\end{eqnarray}}

\newcommand{\Bbar}{\,\overline{\!B}{}}
\newcommand{\Dbar}{\,\overline{\!D}{}}
\newcommand{\Kbar}{\,\overline{\!K}{}}
\def\B0bar{\Bbar{}^0}
\def\D0bar{\Dbar{}^0}
\def\K0bar{\Kbar{}^0}

\begin{document}

\title{\boldmath Glueball-meson molecules}

\def\usc{Department of Physics and Astronomy, 
University of South Carolina, Columbia, SC 29208, USA}
\def\wsu{Department of Physics and Astronomy, 
Wayne State University, Detroit, MI 48201, USA}
\def\mctp{Leinwenber Center for Theoretical Physics, 
University of Michigan, Ann Arbor, MI 48109, USA}
\def\fnal{Theoretical Physics Department, Fermilab, 
P.O. Box 500, Batavia, IL 60510, USA}
\def\munich{Excellence Cluster ORIGINS, 
Technische Universit\"at Mu\"nchen,
Boltzmannstr. 2, D-85748 Garching, Germany}

\author{Alexey A.\ Petrov}
\affiliation{\usc}\affiliation{\wsu}\affiliation{\fnal}\affiliation{\munich}

\preprint{WSU-HEP-2202,FERMILAB-PUB-22-353-T-V}
\date{\today}


\begin{abstract}
Experimental searches for pure glueball states have proven challenging and so far yielded no results. This is believed to occur
because glueballs mix with the ordinary $q\bar q$ states with the same quantum numbers. We will discuss an alternative mechanism, 
the formation of the glueball-meson molecular states. We will argue that the wave functions of already observed excited 
meson states may contain a significant part due to such molecular states. We discuss the phenomenology of glueball-meson 
molecules and comment on a possible charmless component of the $XYZ$ states.
\end{abstract}

\maketitle

\section{Introduction}\label{Intro}

The existence of glueballs, strongly interacting states supposedly made of the pure glue degrees of freedom, is 
expected in Quantum Chromodynamics (QCD). This follows from the non-abelian nature of the theory, where 
the force carriers are also charged under the color gauge group. Various studies of the properties of 
glueballs have been performed both in lattice QCD and in other, more model-dependent approaches. Despite 
various predictions for their properties, no candidates for the pure glueball states have been experimentally 
observed, even in such glue-rich environments as nuclear collisions. Comprehensive reviews of the current state of 
theoretical calculations of glueball properties and their experimental searches can be found 
in \cite{Ochs:2013gi,Klempt:2007cp,Chen:2022asf}.

Spectroscopic studies of glueballs \cite{Ochs:2013gi} suggest that possible quantum numbers of the pure glue 
states, such as $0^{++}, 0^{-+}, 1^{--}$, etc., match with those of the ordinary mesons, i.e., states built out of 
$q \bar q$ pairs. This fact alone implies the quantum mechanical mixing of such states. 

Phenomenological studies of the spectrum of states above one GeV reveal that no resonance that has been experimentally 
discovered so far could be unambiguously classified as a pure glueball state. A consensus appears to be reached 
that the admixture of pure glue and $q\bar q$ states is not tiny, i.e., glueball can only appear as part of the wavefunction 
of ordinary meson states above 1 GeV. Several studies of such mixing have been 
performed \cite{Klempt:2021wpg,Janowski:2011gt,Parganlija:2012fy,Cheng:2015iaa,Frere:2015xxa}. 

In this paper, we propose another possibility. We will prove that the glueballs could also manifest themselves differently 
as glueball-meson molecular states. Depending on the coupling of the glueball and meson states, such molecular states 
may or may not exhibit a universal behavior predicted for the $X(3872)$ or deuterium states. The molecular nature makes 
these states distinct from the hybrid meson states \cite{Juge:2002br}.

For simplicity, here we shall concentrate on the lightest glueball state $G$, the state with the quantum numbers of a 
scalar, $0^{++}$. This is probably the most-studied state with various predictions for its mass and mixing patterns with 
other meson scalar states, all pointing that this is the lightest glueball state with a mass between 1 and 2 GeV.
To determine the properties of the molecular states, we shall employ the so-called extended SU(2) Linear Sigma 
Model (eLSM) \cite{Janowski:2011gt,Parganlija:2012fy}. In this model, the standard SU(2) linear sigma model is complemented 
by the scalar dilaton field $G$, representing the glueball's interpolating field. The parameters of the dilaton-pion
interactions are fixed by fitting the decay widths of the scalar fields, whose masses and mixing parameters are
fitted to the masses of $f_0(1710)$, $f_0(1500)$, and $f_0(1370)$ states. The resulting lowest-energy molecular state will have 
quantum numbers of the pseudoscalar, $0^{-+}$, so we will denote it as ${\cal P}$ in what follows.

Has such a state been observed? We will conjecture that a light quark state $\pi(1800)$ has a significant ${\cal P}$ 
molecular component. This could explain its somewhat narrow width, which is rather unusual for such a massive 
light-quark $q \bar q$ state \cite{EugenioTalk,Barnes:1996ff}. We will also discuss the decay patterns of glueball-meson 
molecular states. The proof of this conjecture would require a thorough investigation of this state's decay and production 
channels. Finally, we will comment on the possible implications of glueball-meson molecules with other quantum numbers.

\section{Sigma model and a molecular state }\label{Model}

To describe pion-glueball interactions, we need to specify a framework in which the 
description of the system would be done. This is especially important for the description of the glueball
states. As it turns out, a symmetry-inspired description of the scalar $0^{++}$ glueball field is possible
and follows from the definition of the QCD trace anomaly. The pure glue sector of classical QCD is
invariant under a dilatation symmetry, which is broken by quantum corrections. This results in the 
non-zero value of the trace of the energy-momentum tensor $T^\mu_\mu$. The vacuum expectation
value of this trace is proportional to the value of the gluon condensate,
\beq
\langle T^\mu_\mu \rangle = 
-\frac{11 N_c}{48} \left\langle\frac{\alpha_s}{\pi} G_{\mu\nu}^a G_a^{\mu\nu} \right\rangle
\equiv -\frac{11 N_c}{48} C^4,
\eeq
where $C^4 = (300-600 \ \mbox{MeV})^4$ \cite{Janowski:2011gt,Parganlija:2012fy}. An effective field theory can 
be built for the scalar field $G$, describing the trace anomaly, where $G$ can be used as an interpolating 
field for the scalar glueball.  

It is convenient to employ the extended linear sigma model, whose parameters have already been 
fitted to describe scalar-glueball mixing and glueball decays \cite{Janowski:2011gt,Parganlija:2012fy}. 
As we shall see, no new interaction terms are needed to describe the molecular ${\cal P}$ states. 
Additional details on the choice of the Lagrangian and the parameter-fitting procedure
can be found in  \cite{Janowski:2011gt,Parganlija:2012fy}.

The effective Lagrangian of the eLSM is
\beq\label{eLSM}
{\cal L} = {\cal L}_{\rm LSM} +  {\cal L}_{\rm dilaton} + {\cal L}_{\rm int},
\eeq
where the first term, ${\cal L}_{\rm LSM}$, corresponds to the Lagrangian of the ordinary 
SU(2) linear sigma model (a generalization to the SU(3) LSM is straightforward \cite{Parganlija:2012fy}),
\bea\label{LSM}
{\cal L}_{\rm LSM} &=&  \mbox{Tr}  \left[\left(\partial^\mu \Phi\right)^\dagger \left(\partial_\mu \Phi\right)
\right] - \lambda_1 \left(\mbox{Tr}  \left[\Phi^\dagger \Phi\right]\right)^2
\nonumber \\
&-& \lambda_2 \ \mbox{Tr}  \left[ \left(\Phi^\dagger \Phi\right)^2\right] 
+  \mbox{Tr}  \left[ H\left(\Phi^\dagger + \Phi\right)\right] 
\\
&+& c\left(\det(\Phi^\dagger) + \det(\Phi) \right),
\nonumber
\eea
where $\Phi$ contains both scalar $\sigma$ and ${a_0}_i$ and pseudoscalar $\eta_N$ and $\pi_i$ fields,
\beq
\Phi = (\sigma + i \eta_N) t^0 + (\vec a_0 + i \vec \pi)\cdot \vec t,
\eeq
where $t^0$ and $t_i$ are the generators of U(2). Note that $\eta_N=(\bar u u + \bar d d)/\sqrt{2}$ only contains 
non-strange degrees of freedom. The last two terms in Eq.~(\ref{LSM}) represent the explicit breaking of chiral symmetry 
$\mbox{Tr}  \left[ H\left(\Phi^\dagger + \Phi\right)\right]  \sim h \sigma$, where $h\sim m_q^2$ 
\cite{Janowski:2011gt,Parganlija:2012fy}, and a contribution due to the chiral anomaly. Those terms also explicitly break 
dilaton symmetry.

The effective Lagrangian for the dilaton field can be written as \cite{Salomone:1980sp,Migdal:1982jp}
\bea\label{Dilaton}
{\cal L}_{\rm dilaton} &=& \frac{1}{2} \left(\partial_\mu \tilde G\right)^2 - \frac{1}{4} \frac{m_G^2}{\Lambda^2} 
\left[\tilde G^4 \log \left|\frac{ \tilde G}{\Lambda}\right| - \frac{1}{4} \tilde G^4
\right]
\nonumber \\
&=& \frac{1}{2} \left(\partial_\mu  \tilde G\right)^2 -  V(\tilde G).
\eea
The minimum of the dilaton potential $V(G)$ is at $ \tilde G = G_0 =\Lambda$. Expanding around the minimum of the 
potential $ \tilde G \to \Lambda + G$ we obtain the effective Lagrangian for the field glueball field $G$
with the mass parameter $m_G \sim 1.6$ GeV obtained from the lattice computations. It can be related 
to the trace of the energy-momentum tensor, implying that \cite{Janowski:2011gt,Parganlija:2012fy}
\beq
\Lambda^2 = \frac{11 C^4}{4 m_G^2}.
\eeq 
The glueball-pion interaction piece ${\cal L}_{\rm int}$ can be written as
\beq\label{int}
{\cal L}_{\rm int} = - m_0^2 \ \mbox{Tr}  \left[\left(\frac{ \tilde G}{\Lambda} \right)^2 \Phi^\dagger \Phi \right]. 
\eeq
This Lagrangian contains both glueball-pion and glueball-sigma interaction terms, which are relevant for 
computing the properties of the molecular state ${\cal P}$. The $\pi \pi G^2$ interaction term can be directly read off 
Eq.~(\ref{int}), while the $\sigma$-exchange interaction can be obtained from the following procedure. Setting all fields 
but $\sigma$ and $G$ in Eq.~(\ref{int}) to zero and shifting the remaining fields by their vacuum expectation values (VEVs)
$\tilde G \to G + G_0$ and $\sigma \to \sigma + \langle \sigma \rangle$ we obtain the $\sigma GG$ interaction term,
\beq\label{Gsigma}
{\cal L}_{\sigma G} = - \frac{m_0^2 \langle \sigma \rangle}{\Lambda^2} G^2 \sigma  + ...
\eeq
The ellipses in Eq.~(\ref{Gsigma}) represent terms that are not relevant for building a molecular state. 
However, they contain terms responsible for mixing glueballs and $q\bar q$ states in this
model. The terms generating the $\pi \pi \sigma$ vertex can also be obtained from Eq.~(\ref{LSM}) by
shifting $\sigma \to \sigma + \langle \sigma \rangle$, as in the usual LSM.

\begin{center}
\begin{figure}
\includegraphics[scale=0.85]{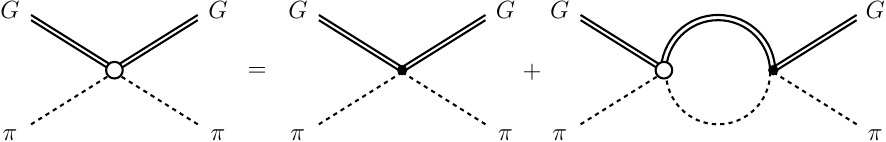}
\caption{Transition amplitude for the $\pi G$ scattering written in the form of a Lippmann-Schwinger equation.  
\label{fig:diagram}}
\end{figure}
\end{center}

The local effective Lagrangian relevant for the scales $\mu < m_\pi$ can be obtained by matching the 
$\pi G$ scattering amplitude and integrating out the $\sigma$ field,
\beq
{\cal L}_{\rm \pi G} = -\lambda \pi^2 G^2,
\eeq
where the effective coupling $\lambda$ is given by 
\beq\label{EffCoupl}
\lambda = \frac{m_0^2}{2 \Lambda^2} \left[1 - \frac{\langle \sigma \rangle^2}{m_\sigma^2}
\left(2\lambda_1 + \lambda_2\right)\right].
\eeq
The numerical values of the parameters in Eq.~(\ref{EffCoupl}) can be obtained from \cite{Parganlija:2012fy}.
To compute the properties of the molecular states ${\cal P}$, we will use the formalism employed in 
describing the properties of deuterium \cite{Weinberg:1990rz} or $X(3872)$ state \cite{AlFiky:2005jd}.
In particular, we will consider non-perturbative scattering amplitude for the scattering $\pi G \to \pi G$. 
The pole of this amplitude, if exists, corresponds to the bound state in the $\pi G$ channel, i.e., the 
molecular ${\cal P}$ state. The transition amplitude can be obtained from the Lippmann-Schwinger 
equation (see Fig.~\ref{fig:diagram}),
\beq\label{LSE}
i T_{\rm \pi G} = - i\lambda + \int \frac{d^4 q}{(2\pi)^4} \left(i T_{\rm \pi G}\right) G_{\rm \pi G} \left(-i\lambda\right)
\eeq
where $G_{\rm \pi G}$ is given by
\bea
G_{\rm \pi G} =  &-& \frac{1}{4 m_G m_\pi} 
\frac{1}{E/2 + q_0 - \vec{q}^2/2 m_G + i \epsilon} 
\nonumber \\
&\times& \frac{1}{E/2 - q_0 - \vec{q}^2/2 m_\pi + i \epsilon}.
\eea
Here $E$ is the energy of the mesons in the center-of-mass frame. The solution of Eq.~(\ref{LSE}) is
\beq
T_{\rm \pi G} = \frac{\lambda}{1+ i \lambda \widetilde A},
\eeq
$\widetilde A$ is the (divergent) integral given by 
\beq\label{WideA}
\widetilde A = -\frac{i}{2} \frac{\mu_{\pi G}}{m_G m_\pi} \int \frac{d^3 q}{(2 \pi)^3} \frac{1}{\vec{q}^2 - 2 \mu_{\pi G} E - i\epsilon},
\eeq
where $\mu_{\pi G}$ is the reduced mass of the $\pi G$ system, and we have used the residue theorem to evaluate the 
integral over $q_0$. 

The integral in Eq.~(\ref{WideA}) diverges, so its divergence needs to be removed by renormalizing the coupling $\lambda$. 
Following S. Weinberg \cite{Weinberg:1990rz,Weinberg:1991um}, we choose to define the renormalized $\lambda_R$ in 
the MS subtraction scheme in dimensional regularization \cite{AlFiky:2005jd}. Computing the integral of Eq.~(\ref{WideA}) in 
$d-1$ dimensions yields
\beq
\widetilde A \to \widetilde A_R = \frac{ip}{8\pi}\frac{\mu_{\pi G}}{m_\pi m_G} ,
\eeq
where $E=p^2/2\mu_{\pi G}$. This implies that the scattering amplitude $T_{\pi G}$  which solves the 
Lippmann-Schwinger equation of Eq.~(\ref{LSE}) is
\beq\label{AmpLSE}
T_{\pi G} = \frac{\lambda_R}{1+ \frac{i\lambda_R}{8\pi} \frac{\mu_{\pi G}}{m_\pi m_G} p}.
\eeq

\noindent
The scattering amplitude of Eq.~(\ref{AmpLSE}) has a pole, which corresponds to the bound state with the energy
\beq\label{BindingE}
E_b =  \frac{32 \pi^2 m_\pi^2 m_G^2}{\lambda_R^2 \mu^3_{\pi G}}.
\eeq
Assuming a nonrelativistic bound state implies that the binding energy is small. The mass of the molecular
state ${\cal P}$ would then be 
\beq
m_{\cal P} = m_G + m_\pi - E_b.
\eeq
Since the bulk of theoretical predictions points to the glueball mass values of 
$m_G \sim 1.6-1.7$ GeV \cite{Ochs:2013gi,Klempt:2007cp,Chen:2022asf},
it is natural to assume that the mass of the molecular state should be around $m_{\cal P} \approx 1.8$ GeV.
Interestingly, the state $\pi(1800)$ has the correct quantum numbers and 
could be seen as a candidate for the glueball-meson molecular state. We will conjecture that ${\cal P}$ 
gives the dominant component of the wave function of this state and discuss some consequences of this 
assumption.

\onecolumngrid
\twocolumngrid

\section{Phenomenology}\label{Pheno}

In the previous section, we predicted the existence of the pseudoscalar glueball-meson molecular state ${\cal P}$ by 
identifying a pole in a $\pi G$ scattering amplitude. Yet, both $\pi$ and $G$ are not
asymptotically stable states. While the small width of the $\pi$, which only decays via weak interactions, can be neglected,
little is known about glueballs' widths. With the glueball decaying strongly, one can legitimately ask if the 
molecular bound state has enough time to form at all. In general, one can argue that in the large $N_c$ limit, the glueball widths
scale like $1/N_c^2$, while the $q\bar q$ meson widths scale only as $1/N_c$ \cite{Nussinov:2009tq,Giacosa:2017eqy}. This gives us 
confidence that the glueball-meson molecular states can be formed. It must be noted that the scalar glueball considered 
in this paper is uniquely linked to the trace anomaly, so its width could be computed \cite{Janowski:2011gt,Parganlija:2012fy}.

It is interesting to consider the phenomenology of the glueball-meson molecular states. One complicating factor, however, exists. 
As we pointed out, quantum mechanics requires that states of a different nature but the same quantum numbers mix. 
If the molecular state mixes with $q\bar q$ or other states, all predictions for the decay patterns will depend on the 
value of the mixing angle(s). We can make predictions for the ratios of various two-body decays {\it assuming} that a 
given physical state is dominated by its glueball-molecular component. If so, the decay patterns of the state
would be driven by the decay patterns of the glueball component. In particular, postulating that ${\cal P}$ is the dominant 
component of the $\pi(1800)$ wave-function, we can determine the ratios of its two-body decays into the $f_0$ states. 

Since the $0^{++}$ glueball's quantum numbers coincide with those of the $f_0$ states, they should contain various 
admixtures of the glue component. The studies of glueball-meson mixing were performed by several authors by studying 
decays and the production of those states. The results of these studies could be summarized in the following matrix equation,
\beq\label{MatrEq1}
\mathbb{F} = \mathbb{M} \ \mathbb{Q},
\eeq
where $\mathbb{F} = \{\left|f_0(1370)\right\rangle,\left|f_0(1500)\right\rangle,\left|f_0(1710)\right\rangle\}$ and 
$\mathbb{Q}=\{\left|\eta_N\right\rangle,\left|\bar s s\right\rangle,\left|G\right\rangle\}$ are the columns for the meson 
and quark/glueball states, respectively. Note that $\eta_N$ was defined in the previous section.

There are several studies that extract the mixing matrix $\mathbb{M}$. One study 
\cite{Cheng:2015iaa,Frere:2015xxa} finds that
\beq\label{M1}
\mathbb{M}_1 = 
\left( 
\begin{array}{ccc}
0.78 & 0.51 & -0.36 \\
-0.54 & 0.84 & -0.03 \\
0.32 & 0.18 & 0.93 
\end{array} 
\right).
\eeq
Alternatively, \cite{Janowski:2011gt,Parganlija:2012fy} give a result for the mixing matrix $\mathbb{M}$ that 
is somewhat similar to the one given in Eq.~(\ref{M1}),
\beq\label{M2}
\mathbb{M}_2 = 
\left( 
\begin{array}{ccc}
0.79 & -0.54 & 0.29 \\
0.49 & 0.84 & 0.22 \\
-0.37 & 0.023 & 0.93 
\end{array} 
\right).
\eeq
It is interesting to note that both studies agree that the only acceptable scenario is that $f_0(1710)$ state is
mostly gluonic, which we used to justify our assumption that the $\pi(1800)$ state is primarily a glueball-meson 
molecule state.

If glueball-molecular component ${\cal P}$ dominates, the decay amplitude $ {\cal A} (\pi(1800) \to \pi f_0)$ can be 
written as
\beq
{\cal A}(\pi(1800) \to \pi f_0) = \langle f_0 | G \rangle \langle \pi G | {\cal H} | \pi(1800)  \rangle,
\eeq
where $\langle f_0 | G \rangle$ parameterizes the glue component of the $f_0$ state for different $f_0$ states,
which can be obtained by inverting the relation in Eq.~(\ref{MatrEq1}), 
\bea\label{GlueComp}
| G \rangle &=& \langle f_0(1370 | G \rangle \left|f_0(1370)\right\rangle 
\\ \nonumber
&+& \langle f_0 (1500) | G \rangle \left|f_0(1500)\right\rangle 
\\
&+& \langle f_0 (1710) | G \rangle \left|f_0(1710)\right\rangle. 
\nonumber 
\eea
Note that other decay mechanisms are OZI-suppressed. 
Then, the ratio of branching ratios for the decays into $\pi f_0(1500)$ and $\pi f_0(1370)$ final states would
be entirely determined by the glue components of the two $f_0$ states, and the difference in phase space
available in each decay,
\beq\label{RatioF}
\frac{{\cal B}(\pi(1800) \to \pi f_0(1500))}{{\cal B}(\pi(1800) \to \pi f_0(1370))} =
\left|\frac{\langle f_0 (1500) | G \rangle}{\langle f_0 (1370) | G \rangle} \right|^2
r_p,
\eeq
where $r_p = p_{f_0(1500)}/p_{f_0(1500)}$ is the ratio of the phase space factors 
for the two-body decay, which is given by the three-momenta of the final state particles, 
\beq
p_{f_0} = \frac{\sqrt{\left(M^2-\left(m_\pi+m_{f_0}\right)^2\right)
\left(M^2-\left(m_\pi-m_{f_0}\right)^2\right)}}{2 M}
\eeq
Numerically, the ratio in Eq.~(\ref{RatioF}) differs slightly depending on 
whether the matrix from Eq.~(\ref{M1}) or (\ref{M2}) is used to obtain Eq.~(\ref{GlueComp}), 
\beq\label{RatioBR}
\frac{{\cal B}(\pi(1800) \to \pi f_0(1500))}{{\cal B}(\pi(1800) \to \pi f_0(1370))} =
(4 \div 7) \times 10^{-3},
\eeq
where the first result corresponds to inverting $\mathbb{M}_1$, while the second - to 
inverting $\mathbb{M}_2$. The trend, however, is similar. That is, assuming that 
the glueball-meson molecule dominates the $\pi(1800)$ wave function, the 
branching ratio of ${\cal B}(\pi(1800) \to \pi f_0(1500))$ is suppressed
compared to ${\cal B}(\pi(1800) \to \pi f_0(1370))$. It is interesting to point out that 
$\pi(1800)$ has been seen in the $\pi f_0(1370)$ decay channel, but not in 
$\pi f_0(1500)$  \cite{ParticleDataGroup:2022pth}, which is consistent 
with the prediction in Eq.~(\ref{RatioBR}). 

\section{Conclusions}\label{Conclusions}

We proposed the existence of a new family of hadronic states, the glueball-meson molecules. The existence of such states 
was justified by showing that the $\pi G$ scattering amplitude contains the pole, whose position
identifies the mass of the state.

Based on the hypothesis that $\pi(1800)$ state contains a dominant ${\cal P}$ glueball molecular component,
we predicted the ratio of decay branching ratios to $\pi f_0(1500)$ and $\pi f_0(1370)$ states. We should note 
that the main decay channels of a glueball-meson molecular state should be driven by the decay of its 
glueball component. In particular, we would expect the ratios of decay widths into $3\pi$, $K\bar K\pi$, and $2 \eta \pi$
final states to mainly follow the quark counting rules for the glueball decays \cite{Ochs:2013gi}.

The molecular states described in this paper would also be different from the hybrid meson states, which can also 
be described in the effective field theory framework \cite{Chiladze:1998ti}. This difference can easiest be seen in the 
context of the flux-tube model of the hybrids, where the gluonic degrees of freedom are encoded in the 
vibrations of the gluonic flux tube connecting the quarks. The meson-glueball molecular state described in this 
paper contains glueball states which are represented by the excitations of the closed flux tubes 
(see e.g., \cite{Iwasaki:2003cr}). This can also be seen from a lattice QCD description of hybrids as 
the states that have a significant overlap with the quark-gluon operators of the type $\bar q \Gamma G q$,
where the quark operators $q$ are in the color-octet configuration. The molecular state described in this
paper involves a color-neutral glueball and a color-neutral pion.

There are some intriguing implications of this proposal. For example, it provides additional insights 
into the puzzle of heavy $XYZ$ states, especially those that decay into charmed mesons. 
A possible (if small) mixture of a vector $1^{- -}$ glueball state $G_V$ with a vector $c\bar c$ state \cite{Suzuki:2002bz} 
could provide a glueball-meson molecular component $\pi G_V$ of the $X(3872)$ state containing no charmed quarks at all.  
This follows if the mass of the vector glueball is $m_{G_V} \approx 3.8$ GeV, as suggested by some lattice
QCD studies \cite{Chen:2022asf}. Other implications would include the existence of the glueball-meson molecular 
states with heavy (charm or bottom) quarks. We will address these implications in a forthcoming publication.

\section*{Acknowledgments}

The author would like to thank Ted Barnes, Jozef Dudek, Fred Myhrer, and Eric Swanson for their valuable comments.
This research was supported in part by the U.S.\ Department of Energy under contract DE-SC0007983
and by the Excellence Cluster ORIGINS, which is funded by the Deutsche Forschungsgemeinschaft 
(DFG, German Research Foundation) under Germany's Excellence Strategy - EXC-2094 - 390783311.
It was also supported by the Visiting Scholars Award Program of the Universities Research Association. 
Fermilab is managed and operated by Fermi Research Alliance, LLC under Contract No. DE-AC02-07CH11359 
with the U.S. Department of Energy.


\end{document}